\documentclass[[12pt,a4paper,preprint,preprintnumbers,amsmath,amssymb,nofootinbib]{revtex4}
\usepackage{graphicx}
\usepackage{amsmath}
\usepackage{amssymb}
\usepackage{bbold}
\usepackage{type1cm}
\usepackage{rotating}
\usepackage{tabularx}
\usepackage{pdflscape}

\usepackage{dsfont}
\usepackage{overpic}
\usepackage{natbib}
\usepackage{slashed}
\usepackage{subfigure}
\usepackage{bbm}

\begin{document}
\title{Structure of resonances in a square well potential}

\author{Peter C.~Bruns}
\affiliation{Nuclear Physics Institute, 25068 \v{R}e\v{z}, Czech Republic }
\date{\today}
\begin{abstract}
We study the structure of resonances as derived from the exactly solvable Lippmann-Schwinger equation for a one-dimensional square well potential. Within this framework, we discuss the concept of resonance form factors, and the relation of the corresponding spatial densities to ``resonance wave functions''. 
\end{abstract}

\maketitle

\section{Introduction}
\label{sec:Intro}

In a recent preprint \cite{Bruns:2018huz}, the concept of ``resonance form factors'' was examined in the framework of an exactly solvable Lippmann-Schwinger equation (LSE) for a simple model potential in one space-dimension. It was found that the natural extension of the form factor concept from bound states to resonances leads to certain associated spatial densities $D_{R}(x)$ replacing the absolute squares of the bound-state wave functions. In the present contribution, we would like to exemplify these issues employing a more common potential, namely the ``square well'' potential often encountered in introductory courses on quantum theory (see e.g. Chapter 5 of \cite{Hall}). Our motivation mainly comes from hadron physics. Since most hadrons are unstable particles which can show up as resonances in various processes, it is desirable to deepen our understanding of what we can say about the structure and the ``nature'' of resonances (e.g. quark-antiquark or three-quark states vs. ``hadronic molecules'', etc.). We refer to \cite{Guo:2017jvc} for a recent review of these matters. Intuitively, one thinks that we should be able to determine form factors of sufficiently long-lived unstable particles (see \cite{Sekihara:2008qk,Bauer:2012at,Djukanovic:2013mka} for examples in the field of low-energy hadron physics), and subsequently learn something about the pertaining spatial charge distributions, possibly in the sense of a ``snapshot'' of the distributions sufficiently long after the formation, and long before the decay, of the unstable state. However, the resonance analogues of bound-state wave functions usually encountered in quantum-mechanical treatments of resonances \cite{Gamow:1928zz,Zeldovich:1961a,Berggren:1968zz,Garcia-Calderon:1976omn} are non-normalizable, and one wonders how exactly such (generalized) spatial probability densities can be related to e.g. finite mean square radii of form factors. We hope that the present work will contribute to the clarification of these questions. \\
This article is organized as follows: In Sec.~\ref{sec:LS1d}, we give an outline of the LSE framework, to fix our notation and conventions. This section is essentially copied from \cite{Bruns:2018huz}, just to make the present follow-up article self-contained. In Sec.~\ref{sec:LSpotwell}, we present the solution to the LSE for the case of the square well potential. The main part of this contribution, Sec.~\ref{sec:FF}, starts with the calculation of the form factors related to the probability density $\psi^{\ast}_{B}(x)\psi_{B}(x)$ for the bound states $B$, and then explains a suggested generalization to resonances $R$. It also contains some illustrative examples for suitably chosen parameters of the potential. We summarize our findings in Sec.~\ref{sec:conclusio}. Some additional material, useful for a more detailed analysis of our results, is presented in the appendices.  


\section{Lippmann-Schwinger equation: Generalities}
\label{sec:LS1d}

In this work, we consider non-relativistic quantum mechanics on a one-dimensional space. The Schr\"odinger equations are of the form $\hat{H}|\psi_{n}\rangle = E_{n}|\psi_{n}\rangle\,$, $\hat{H}=\hat{H}_{0}+\hat{V}\,$, $\hat{H}_{0}|p\rangle = E_{n}^{(0)}|p\rangle=\frac{|p|^2}{2\mu}|p\rangle\,$, where $\hat{H}$ is the full Hamilton operator, while $\hat{H}_{0}$ is its ``free'' part. Throughout, we shall use hats over a symbol to indicate its operator character, and deal only with a local potential $\hat{V}$, i.e. $\langle y|\hat{V}|x\rangle = V(x)\delta(x-y)\,$. We define resolvent functions (operators) $\hat{G}_{0}(E)=(E-\hat{H}_{0})^{-1}\,,\quad \hat{G}(E)=(E-\hat{H})^{-1}\,$ for the free and the full Hamilton operator, respectively. 
Expressing $\hat{H}$ as a differential operator in coordinate space, $G(x',x):=\langle x'|\hat{G}(E)|x\rangle$ solves
\begin{equation}
(E-\hat{H})G(x',x) = \delta(x'-x) = \sum_{n}\psi_{n}(x')\psi^{\ast}_{n}(x)\,,\qquad \langle x|\psi_{n}\rangle =: \psi_{n}(x)\,,
\end{equation}
invoking the completeness relation for an orthonormal set of eigenfunctions of $\hat{H}$ (the summation is to be replaced by integration for the continuous part of the spectrum). In operator form, we can express $\hat{G}$ more generally as
\begin{equation}
\hat{G}(E) = \sum_{n}\frac{|\psi_{n}\rangle\langle\psi_{n}|}{E-E_{n}}\,.\qquad\mathrm{Similarly,}\quad \hat{G}_{0}(E) = \int dp\frac{|p\rangle\langle p|}{E-\frac{p^2}{2\mu}}\,.
\end{equation}
We normalize the eigenstates of $\hat{H}_{0}$ as $\langle p'|p\rangle = \delta(p\,'-p)$, so that
\begin{equation}
\langle x|p\rangle = \frac{e^{ipx}}{\sqrt{2\pi}}\qquad\mathrm{and}\qquad \sqrt{2\pi}\langle p|\psi_{n}\rangle =\int dx \,e^{-ipx}\psi_{n}(x)=:\tilde{\psi}_{n}(p)\,,
\end{equation}
employing  units where $\hbar=1$. The Lippmann-Schwinger equation (LSE) reads (see e.g. \cite{Taylor})
\begin{equation}\label{eq:LSE}
\hat{\mathcal{T}}(E)=\hat{V}+\hat{\mathcal{T}}(E)\hat{G}_{0}(E)\hat{V}\,.
\end{equation}
One can convince oneself that the solution can also be written in terms of the full resolvent $\hat{G}$,
\begin{equation}\label{eq:LSG}
  \hat{\mathcal{T}}(E)=\hat{V}+\hat{V}\hat{G}(E)\hat{V}\,,
\end{equation}
since $\hat{G}=\hat{G}_{0}+\hat{G}\hat{V}\hat{G}_{0}=\hat{G}_{0}+\hat{G}_{0}\hat{V}\hat{G}$. Taking matrix elements of Eq.~(\ref{eq:LSE}), we get
\begin{equation}
\langle q'|\hat{\mathcal{T}}(E)|q\rangle = \langle q'|\hat{V}|q\rangle + \int dl\,\frac{\langle q'|\hat{\mathcal{T}}(E)|l\rangle\langle l|\hat{V}|q\rangle}{E-\frac{l^2}{2\mu}}\,.
\end{equation}
 On the other hand, taking matrix elements of (\ref{eq:LSG}),
\begin{eqnarray}
  \langle q'|\hat{\mathcal{T}}(E)|q\rangle &=& \langle q'|\hat{V}|q\rangle + \sum_{n}\frac{\langle q'|\hat{V}|\psi_{n}\rangle\langle\psi_{n}|\hat{V}|q\rangle}{E-E_{n}}\nonumber \\
  &=& \langle q'|\hat{V}|q\rangle + \sum_{n}\frac{\left(\frac{q'^2}{2\mu}-E_{n}\right)\tilde{\psi}_{n}(q')\tilde{\psi}_{n}^{\ast}(q)\left(\frac{q^2}{2\mu}-E_{n}\right)}{2\pi(E-E_{n})}\,,\label{eq:LSGmat}
\end{eqnarray}
by means of the Schr\"odinger equation $\langle q|\hat{V}|\psi_{n}\rangle=-\langle q|\hat{H}_{0}-E_{n}|\psi_{n}\rangle\,$ and the completeness relations. The matrix element $\langle q'|\hat{\mathcal{T}}(E)|q\rangle\equiv \mathcal{T}(q',q;E)$ is the off-shell scattering amplitude for incoming and outgoing momentum $q$ and $q'$, respectively, and energy $E$. \,-\, The integral 
%
\begin{equation}\label{eq:LSEloop}
I_{0}(E):=\int_{-\infty}^{\infty}\frac{dl}{\frac{l^2}{2\mu}-E} = \frac{2\pi i\mu}{k(E)}\,,\qquad k(E)=+\sqrt{2\mu E}\,,
\end{equation}
plays a special role in the LSE framework. Considered as a function of the energy $E$, it has a branch cut along $E>0$ in the complex $E$-plane. Real positive values of $E$ are to be approached from the upper complex plane, and $k(E)=+\sqrt{2\mu E}$ is the positive square root; otherwise, $k(E)\equiv k$ is fixed to be the square root with the positive imaginary part. If not stated otherwise, the variable $k$ will always stand for this square root. These requirements define the first or ``physical'' Riemann sheet. Considering the integral in Eq.~(\ref{eq:LSEloop}) as a function of complex $k$, and continuing analytically to the lower complex $k$-plane, amounts to the analytic continuation in the variable $E$ over the positive real $E$-axis, to the second Riemann sheet. These analytic properties of $I_{0}(E)$ will also show up in the solution of the LSE. The relation between $I_{0}(E)$ and the scalar relativistic two-point loop integral is explained in App.~B of \cite{Bruns:2018huz}.


\section{LSE for the square well potential}
\label{sec:LSpotwell}

For a one-dimensional potential well of length $2d>0$,
\begin{equation}
V(x) = \theta(d-x)\theta(x+d)V_{0}\quad\Rightarrow\quad \langle q'|\hat{V}|q\rangle = \int_{-d}^{d}\frac{dx}{2\pi}\,e^{-i(q'-q)x}V_{0} = \frac{V_{0}d}{\pi}\left(\frac{\sin((q'-q)d)}{(q'-q)d}\right)\,,
\end{equation}
where $\theta(\cdot)$ denotes the Heaviside step function, one finds the following solution to the LSE:
\begin{eqnarray}
  \mathcal{T}(q',q;E) &=& \mathcal{T}_{\mathrm{sep}}(q',q;E) + \mathcal{T}_{\mathrm{bg}}(q',q;E)\,,\label{eq:fullsolT}\\
  \mathcal{T}_{\mathrm{sep}}(q',q;E) &=& \frac{w_{0}(E)}{q'^2-\xi^2}\biggl((q'-k)(q-k)e^{i(q'+q)d} + (q'+k)(q+k)e^{-i(q'+q)d} \nonumber \\ &+& e^{2i\xi d}\frac{k-\xi}{k+\xi}\biggl((q'-k)(q+k)e^{i(q'-q)d} + (q'+k)(q-k)e^{-i(q'-q)d}\biggr)\biggr)\frac{1}{q^2-\xi^2}\,,\nonumber \\
  \mathcal{T}_{\mathrm{bg}}(q',q;E) &=& \frac{V_{0}}{2\pi i(q'-q)}\biggl(e^{i(q'-q)d}\left(\frac{q'-k}{q'-\xi}\right)\left(\frac{q+k}{q+\xi}\right) - e^{-i(q'-q)d}\left(\frac{q'+k}{q'+\xi}\right)\left(\frac{q-k}{q-\xi}\right)\biggr)\,,\nonumber \\
  w_{0}(E) &=& \frac{4i\mu\xi V_{0}^2 e^{2i\xi d}}{2\pi\left((k+\xi)^2-e^{4i\xi d}(k-\xi)^2\right)}\,, \quad k=+\sqrt{2\mu E}\,,\quad \xi=+\sqrt{2\mu(E-V_{0})}\,. \nonumber 
\end{eqnarray}
We see that the solution can be split in a ``separable'' part $\mathcal{T}_{\mathrm{sep}}$, which can be written as a finite sum of products $\sim\sum_{n} f_{n}(q')g_{n}(q)$\,, and contains all the bound-state and resonance poles encoded in the function $w_{0}(E)$, and a ``background'' amplitude $\mathcal{T}_{\mathrm{bg}}$ which is free of poles in $E$ for on-shell momenta ($|q'|=|q|=k$).\\ 
We can define the analogue of on-shell s-and p-wave ``partial wave'' amplitudes (strictly speaking, scattering amplitudes for states of positive/negative parity) by fixing $q=k$ and averaging over $q'=\pm k$,
\begin{equation}
\mathcal{T}_{\mathrm{s}}(E):= \frac{1}{2}\left(\mathcal{T}(+k,k;E)+\mathcal{T}(-k,k;E)\right)\,,\quad \mathcal{T}_{\mathrm{p}}(E):= \frac{1}{2}\left(\mathcal{T}(+k,k;E)-\mathcal{T}(-k,k;E)\right)\,.\label{eq:pwdef}
\end{equation}
They can be written in the ``K-Matrix'' form
\begin{eqnarray}
  \mathcal{T}_{\mathrm{s}}(E) &=& \left\lbrack K_{\mathrm{s}}^{-1}+\frac{2\pi i\mu}{k}\right\rbrack^{-1}\,,\quad \mathcal{T}_{\mathrm{p}}(E) = \left\lbrack K_{\mathrm{p}}^{-1}+\frac{2\pi i\mu}{k}\right\rbrack^{-1}\,,\label{eq:pwKform}\\
  K_{\mathrm{s}} &=& \frac{k}{2\pi\mu}\left(\frac{(k+\xi)\sin((k-\xi)d)+(k-\xi)\sin((k+\xi)d)}{(k+\xi)\cos((k-\xi)d)+(k-\xi)\cos((k+\xi)d)}\right)\,,\label{eq:Ks}\\
  K_{\mathrm{p}} &=& \frac{k}{2\pi\mu}\left(\frac{(k+\xi)\sin((k-\xi)d)-(k-\xi)\sin((k+\xi)d)}{(k+\xi)\cos((k-\xi)d)-(k-\xi)\cos((k+\xi)d)}\right)\,.\label{eq:Kp}
\end{eqnarray}
The functions $K_{\mathrm{s,p}}(E)\equiv K_{\mathrm{s,p}}$ play the role of ``effective potentials'' here. Note that they are even in $k$ and in $\xi$, so they are real for real $E$, and do not possess branch cuts.\\ 
The parity-even amplitude $\mathcal{T}_{\mathrm{s}}(E)$ has poles for
\begin{equation}
k+\xi + e^{2i\xi d}(k-\xi) =0\,, \quad\mathrm{or}\quad \tan(\xi d) = \frac{\kappa}{\xi}\,,\label{eq:polcondplus}
\end{equation}
while the parity-odd amplitude $\mathcal{T}_{\mathrm{p}}(E)$ has poles for
\begin{equation}
k+\xi - e^{2i\xi d}(k-\xi) =0\,, \quad\mathrm{or}\quad \tan(\xi d) = -\frac{\xi}{\kappa}\,,\label{eq:polcondminus}
\end{equation}
where we set $k=i\kappa$. For a bound state of energy $E_{B}<0$, we have $+\sqrt{2\mu(-E_{B})}=\kappa_{B}=-ik_{B}>0$. The resonance poles are located on the second Riemann sheet in $E$, and have $\mathrm{Im}\,k_{R}<0$. Let there be bound-state solutions for $(k_{B}^{\pm}=i\kappa_{B}^{\pm},\,\xi_{B}^{\pm})$ with energy $E_{B}^{\pm}$, i.e.
\begin{displaymath}
(k_{B}^{\pm} + \xi_{B}^{\pm}) \pm (k_{B}^{\pm} - \xi_{B}^{\pm})e^{2i\xi_{B}^{\pm} d}=0\,.
\end{displaymath}
For an energy $E$ close to the respective bound-state energy, the off-shell amplitude behaves as
\begin{eqnarray*}
  \mathcal{T}(q',q;E\rightarrow E_{B}^{+}) &\,\rightarrow\,& \frac{4V_{0}^2(\xi_{B}^{+})^2\kappa_{B}^{+}\left(q'\sin(q'd)-\kappa_{B}^{+}\cos(q'd)\right)\left(q\sin(qd)-\kappa_{B}^{+}\cos(qd)\right)}{2\pi(E-E_{B}^{+})(q'^2-(\xi_{B}^{+})^2)((\kappa_{B}^{+})^2+(\xi_{B}^{+})^2)(1+\kappa_{B}^{+}d)(q^2-(\xi_{B}^{+})^2)}\,,\\
  \mathcal{T}(q',q;E\rightarrow E_{B}^{-}) &\,\rightarrow\,& \frac{4V_{0}^2(\xi_{B}^{-})^2\kappa_{B}^{-}\left(q'\cos(q'd)+\kappa_{B}^{-}\sin(q'd)\right)\left(q\cos(qd)+\kappa_{B}^{-}\sin(qd)\right)}{2\pi(E-E_{B}^{-})(q'^2-(\xi_{B}^{-})^2)((\kappa_{B}^{-})^2+(\xi_{B}^{-})^2)(1+\kappa_{B}^{-}d)(q^2-(\xi_{B}^{-})^2)}\,.
\end{eqnarray*}
Employing the results of App.~\ref{app:LSpotwellwf}, in particular Eqs.~(\ref{eq:SchrBpotplus}), (\ref{eq:SchrBpotminus}), this can be rewritten in terms of the bound-state wave functions,
\begin{equation}
\mathcal{T}(q',q;E\rightarrow E_{B}^{\pm}) \quad\rightarrow\quad \frac{\left(\frac{q'^2}{2\mu}-E_{B}^{\pm}\right)\tilde{\psi}^{\pm}_{B}(q')\tilde{\psi}_{B}^{\pm\ast}(q)\left(\frac{q^2}{2\mu}-E_{B}^{\pm}\right)}{2\pi(E-E_{B}^{\pm})}\,.\label{eq:reswfpot}
\end{equation}
This establishes the relation between the residues of $\mathcal{T}$ at the bound-state poles and the bound-state wave functions, in accord with Eq.~(\ref{eq:LSGmat}).

\section{Form factors}
\label{sec:FF}

Given the bound state wave functions of the previous section, we can define the form factor of a bound state $B$ as
\begin{eqnarray}
  F_{B}(Q^2) &:=& \int_{-\infty}^{\infty}dx\,\psi^{\ast}_{B}(x)e^{iQx}\psi_{B}(x) = 1-\frac{1}{2}Q^2\langle\hat{x}^2\rangle_{B} + \mathcal{O}(Q^4)\,,\label{eq:defF}\\
  \langle\hat{x}^2\rangle_{B} &:=& \int_{-\infty}^{\infty}dx\,\psi^{\ast}_{B}(x)x^2\psi_{B}(x)\,.\label{eq:defxsqr}
\end{eqnarray}
In terms of the momentum-space wave functions, the form factor can be rewritten as
\begin{eqnarray}
  F_{B}(Q^2) &=& \int_{-\infty}^{\infty}\frac{dp}{2\pi}\,\tilde{\psi}^{\ast}_{B}(p+Q)\tilde{\psi}_{B}(p) \nonumber \\
  &=& \int_{-\infty}^{\infty}\frac{dp}{2\pi}\,\frac{\tilde{\psi}^{\ast}_{B}(p+Q)\left(\frac{(p+Q)^2}{2\mu}-E_{B}\right)\left(\frac{p^2}{2\mu}-E_{B}\right)\tilde{\psi}_{B}(p)}{\left[\frac{(p+Q)^2}{2\mu}-E_{B}\right]\left[\frac{p^2}{2\mu}-E_{B}\right]}\,.
\end{eqnarray}
Making use of Eq.~(\ref{eq:reswfpot}), we can rewrite this further in terms of the residuum of the scattering amplitude at the bound-state pole:
\begin{equation}
  F_{B}(Q^2) = \frac{1}{\mathrm{Res}\,\mathcal{T}(q',q;E\rightarrow E_{B})}\int_{-\infty}^{\infty}dp\,\frac{\left(\mathrm{Res}\,\mathcal{T}(q',p+Q;E\rightarrow E_{B})\right)\left(\mathrm{Res}\,\mathcal{T}(p,q;E\rightarrow E_{B})\right)}{\left[\frac{(p+Q)^2}{2\mu}-E_{B}\right]\left[\frac{p^2}{2\mu}-E_{B}\right]}\,.\label{eq:FFgeneral}
\end{equation}
The integral
\begin{equation}\label{eq:F0}
\mathcal{F}_{0}(Q^2;E):=\int_{-\infty}^{\infty}\frac{dp}{\left[\frac{(p+Q)^2}{2\mu}-E\right]\left[\frac{p^2}{2\mu}-E\right]} = \frac{8\pi i\mu^2}{k(Q^2-4k^2)}\,
\end{equation}
can be related to the non-relativistic limit of the relativistic loop integral with three propagators in $1+1$ dimensions, compare App.~B of \cite{Bruns:2018huz}. \\
For the $\pm$ bound states, we compute the form factors, in one way or the other, to be
\begin{equation}\label{eq:FBresult}
F_{B}^{(\pm)}(Q^2) = \frac{4\mathcal{N}_{B}^2((\kappa_{B}^{\pm})^2+(\xi_{B}^{\pm})^2)e^{-2\kappa^{\pm}_{B}d}}{(4(\xi_{B}^{\pm})^2-Q^2)(4(\kappa_{B}^{\pm})^2+Q^2)}\left(4\kappa_{B}^{\pm}\cos(Qd)+(4(\kappa_{B}^{\pm})^2-Q^2)\frac{\sin(Qd)}{Q}\right),
\end{equation}
see Eq.~(\ref{eq:psiBadd}) for the normalization factor $\mathcal{N}_{B}$. Written in this form, the expression is the same for the $(+)$ and $(-)$ solution, but the relation between $\kappa_{B}^{\pm}$ and $\xi_{B}^{\pm}$ is different in each case (Eqs.~(\ref{eq:polcondplus}), (\ref{eq:polcondminus})). For the ``mean square radii'', we find
\begin{equation}\label{eq:xsqrB}
\langle\hat{x}^2\rangle_{B}^{(\pm)} = \frac{d^2}{3} + \frac{1}{2(\kappa_{B}^{\pm})^2} - \frac{1}{2(\xi_{B}^{\pm})^2} + \frac{d(3+4\kappa_{B}^{\pm}d)}{6\kappa_{B}^{\pm}(1+\kappa_{B}^{\pm}d)}\,.
\end{equation}
As a check, one can verify that the inverse Fourier transforms of these form factors yield the absolute squares of the corresponding bound-state wave functions of Eqs.~(\ref{eq:psiBp}),\,(\ref{eq:psiBm}),
\begin{equation}\label{eq:psiBcheck}
\int_{-\infty}^{\infty}\frac{dQ}{2\pi}\,e^{-iQx}F_{B}^{(\pm)}(Q^2) = \psi^{\pm\ast}_{B}(x)\psi_{B}^{\pm}(x)\,.
\end{equation}
A generalization to the case of resonances $R$ (instead of bound states $B$) is suggested by Eq.~(\ref{eq:FFgeneral}). Namely, we consider the amplitude $\mathcal{M}$ for the scattering process in the presence of an external source\footnote{In the present case, it is taken as a scalar source field. A possible coupling constant with which the field couples to the scattering particles is assumed to be divided out in our results for the amplitudes $\mathcal{M}$.}, which transfers a momentum $\sim Q$ to the system,
\begin{equation}\label{eq:Mfull}
\mathcal{M}(q',q;Q,E) := \langle q'|\hat{G}_{0}^{-1}(E)\hat{G}(E)e^{iQ\hat{x}}\hat{G}(E)\hat{G}_{0}^{-1}(E)|q\rangle\,.
\end{equation}
The part of $\mathcal{M}$ which contains the double pole term $\sim(E-E_{R})^{-2}$ for a given resonance $R$ is 
\begin{equation}\label{eq:Mdp}
\mathcal{M}_{\mathrm{d.p.}}(q',q;Q,E) = \int_{-\infty}^{\infty}dp\,\frac{\mathcal{T}(q',p+Q;E)\mathcal{T}(p,q;E)}{\left[\frac{(p+Q)^2}{2\mu}-E\right]\left[\frac{p^2}{2\mu}-E\right]}\,.
\end{equation}
\begin{figure}[t]
\centering
\includegraphics[width=0.45\textwidth]{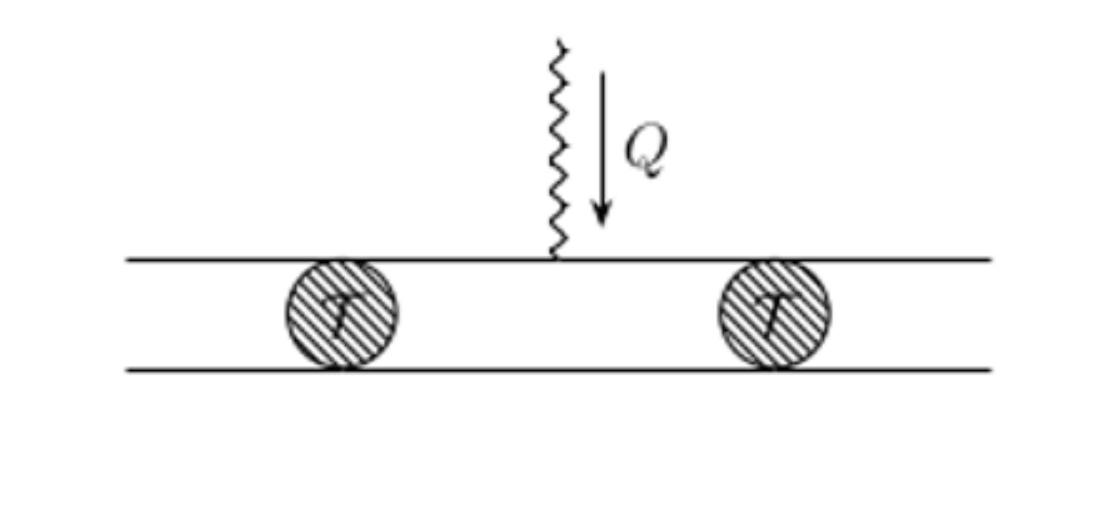}
\caption{Illustration of $\mathcal{M}_{\mathrm{d.p.}}$, where $\mathcal{T}$ appears as an effective vertex. The straight lines stand for the scattering particles, while the wiggly line indicates the external source field. Note that, in the present context of potential scattering, the two involved particles are described as a one-particle system, with reduced mass $\mu$.}
\label{fig:MdpPic}
\end{figure}%
Compare also the ``Feynman graph'' for $\mathcal{M}_{\mathrm{d.p.}}$ in Fig.~\ref{fig:MdpPic}, where time runs horizontally.\\
The expression on the r.h.s. of Eq.~(\ref{eq:F0}) makes it plausible that $\mathcal{M}_{\mathrm{d.p.}}$ has the same cut structure in the complex $E$-plane as the scattering amplitude $\mathcal{T}$. Therefore, we should be able to analytically continue $\mathcal{M}_{\mathrm{d.p.}}$ to the second Riemann sheet, where the resonance pole $E_{R}$ is located, and {\em define\,} the ``resonance form factor'' as
\begin{equation}\label{eq:defFR}
F_{R}(Q^2) := \frac{\lim_{E\rightarrow E_{R}}\left((E-E_{R})^2\mathcal{M}_{\mathrm{d.p.}}(q',q;Q,E)\right)}{\mathrm{Res}\,\mathcal{T}(q',q;E_{R})}\,.
\end{equation}
We refer to \cite{Gegelia:2009py,Hoja:2010fm,Bernard:2012bi} for a further discussion of such a definition in a quantum-field-theoretical context. Note that we could have substituted $\mathcal{T}_{\mathrm{sep}}$ for $\mathcal{T}$ in Eq.~(\ref{eq:defFR}), since the non-separable part $\mathcal{T}_{\mathrm{bg}}$ does not contain the resonance poles, and so the dependence on $q',q$ drops out in $F_{R}$. \\
\quad\\
Let us check the validity of the above reasoning explicitly for our example. 
First we note that, from Eqs.~(\ref{eq:fullsolT}) and (\ref{eq:Mdp}), we have $\mathcal{M}_{\mathrm{d.p.}}(-q',-q;Q,E)=\mathcal{M}_{\mathrm{d.p.}}(q',q;-Q,E)$ (parity). So the full on-shell information of $\mathcal{M}_{\mathrm{d.p.}}$ is contained in the two functions $\mathcal{M}_{\mathrm{d.p.}}(k,k;Q,E)$ and $\mathcal{M}_{\mathrm{d.p.}}(-k,k;Q,E)$\,, evaluated on the physical sheet ($\mathrm{Im}\,k>0$). We give explicit expressions for the latter two functions in App.~\ref{app:Mdp}, in a form which makes it easy to check that they are invariant under the formal transformation $\xi\rightarrow -\xi$, and otherwise meromorphic in $k$ for any fixed $Q$. So we can confirm that $\mathcal{M}_{\mathrm{d.p.}}$ has the same Riemann sheet structure in $E$ as $\mathcal{T}$, and consider the proper analytic continuations of $\mathcal{T}$ {\em and\,} $\mathcal{M}_{\mathrm{d.p.}}$ to the second Riemann sheet (in practice, we might tacitly consider them as functions of $k$, with $\xi^2=k^2-2\mu V_{0}$, and continue our formulas to $\mathrm{Im}\,k<0$). Arriving at a resonance pole, formula (\ref{eq:defFR}) can be applied, and it is straightforward to find ($k_{R}^2=2\mu E_{R}$, $\mathrm{Im}\,k_{R}<0$):
\begin{equation}\label{eq:FRresult}
F_{R}^{(\pm)}(Q^2) = \frac{4k_{R}^{\pm}(\xi_{R}^{\pm})^2\left(4k_{R}^{\pm}\cos(Qd)-i(4(k_{R}^{\pm})^2+Q^2)\frac{\sin(Qd)}{Q}\right)}{(4(\xi_{R}^{\pm})^2-Q^2)(4(k_{R}^{\pm})^2-Q^2)(1-ik_{R}^{\pm}d)}\,,
\end{equation}
which is given by the analytic continuation of the result in Eq.~(\ref{eq:FBresult}) in $E$, as one might have expected from the start (this was also used in \cite{Bruns:2018huz}). In particular, we find the same result for $(q',q)=(k,k)$ and $(-k,k)$ in (\ref{eq:defFR}), so that the on-shell amplitudes indeed give an unambiguous result for $F_{R}^{\pm}$ (for simple separable potentials $\sim v(q')v(q)$, this would be trivially fulfilled).  \\
%
%
\quad \\
We are now looking for those ``densities'' that generate the above form factors via Fourier transformation, in analogy to Eq.~(\ref{eq:defF}). Evaluating the inverse Fourier transforms, taking into account $\mathrm{Im}\,k_{R}^{\pm}<0$ and the pole conditions (\ref{eq:polcondplus}), (\ref{eq:polcondminus}), our result is
\newpage
\begin{eqnarray}
  D_{R}^{(\pm)}(x) &:=& \int_{-\infty}^{\infty}\frac{dQ}{2\pi}\,e^{-iQx}F_{R}^{(\pm)}(Q^2) \label{eq:DRpm} \\
  &=& \mathcal{N}_{R}^2\biggl(\theta(-d-x)e^{2ik_{R}^{\pm}x} + \theta(x-d)e^{-2ik_{R}^{\pm}x} \nonumber \\ &+& \theta(d-x)\theta(x+d)\left(2\cos(2k_{R}^{\pm}x)+\frac{e^{2ik_{R}^{\pm}d}}{2(\xi_{R}^{\pm})^2}\left((k_{R}^{\pm})^2-(\xi_{R}^{\pm})^2\right)\left(1\pm \cos(2\xi_{R}^{\pm}x)\right)\right)\biggr)\,,  \nonumber \\
  \mathcal{N}_{R}^2 &=& \frac{k_{R}^{\pm}(\xi_{R}^{\pm})^2e^{-2ik_{R}^{\pm}d}}{((k_{R}^{\pm})^2-(\xi_{R}^{\pm})^2)(k_{R}^{\pm}d+i)}\,.\label{eq:NR}
\end{eqnarray}
In the following, we discuss the relation between the above densities and the squares of the usual ``resonance wave functions'' encountered in the literature \cite{Gamow:1928zz,Zeldovich:1961a,Berggren:1968zz,Garcia-Calderon:1976omn} - in our case,
\begin{eqnarray}
  \psi_{G}^{+}(x) &=& \mathcal{N}_{G}\biggl(\theta(-d-x)e^{-ik_{R}^{+} x} + \theta(d-x)\theta(x+d)e^{ik_{R}^{+}d}\frac{\cos\xi_{R}^{+} x}{\cos\xi_{R}^{+} d} + \theta(x-d)e^{ik_{R}^{+} x}\biggr)\,,\label{eq:psiGp}\\
  \psi_{G}^{-}(x) &=& \mathcal{N}_{G}\biggl(\theta(-d-x)e^{-ik_{R}^{-} x} - \theta(d-x)\theta(x+d)e^{ik_{R}^{-}d}\frac{\sin\xi_{R}^{-} x}{\sin\xi_{R}^{-} d} - \theta(x-d)e^{ik_{R}^{-} x}\biggr)\,,\label{eq:psiGm}
 \end{eqnarray}
where $\mathcal{N}_{G}^2=-\mathcal{N}_{R}^2$, and the $G$ stands for ``Gamow'' \cite{Gamow:1928zz}. The $\psi_{G}^{\pm}$ are given\footnote{Formally, they can be derived as follows. From our solution to the LSE, we can compute the full Green function for the square well potential via $\hat{G}=\hat{G}_{0}+\hat{G}_{0}\hat{\mathcal{T}}\hat{G}_{0}$. It inherits the appropriate analytic properties from $\mathcal{T}$. Continuing $\langle x'|\hat{G}|x\rangle \equiv G(x',x)$ to the second Riemann sheet, one finds that \begin{displaymath} (E-E_{R}^{\pm})G(x',x)\,\rightarrow\, \psi_{G}^{\pm}(x')\psi_{G}^{\pm}(x) \end{displaymath} (no complex conjugation on the r.h.s.) when $E$ approaches a resonance pole $E_{R}^{\pm}$ associated with positive/negative parity. Note that the $\psi_{G}^{\pm}(x)$ are not normalizable in the standard way, since $\mathrm{Im}\,k_{R}^{\pm}<0$.} by the analytic continuation of the bound-state wave functions of Eqs.~(\ref{eq:psiBp}), (\ref{eq:psiBm}). Consequently, we find the result $(\psi_{G}^{\pm}(x))^2$ when we compute the inverse Fourier transform appearing in Eqs.~(\ref{eq:psiBcheck}), (\ref{eq:DRpm}) for a {\em bound state\,} ($\mathrm{Im}\,k_{B}^{\pm}>0$), and {\em then\,} continue the obtained formula to $k_{R}^{\pm}$ on the second complex-energy sheet, instead of calculating the inverse Fourier transform directly for $k_{R}^{\pm}$. Writing the form factor more generally as
\begin{equation}
F(Q^2) = \frac{\bar{f}(Q)}{Q^2-4k_{R}^2}\,,
\end{equation}
where $\bar{f}(Q)$ is an even function without poles in $Q$, the difference between the outcomes of the two procedures is given by
\begin{equation}
(\psi_{G}^{\pm}(x))^2-D_{R}^{(\pm)}(x) = \frac{i\bar{f}(2k_{R}^{\pm})}{2k_{R}^{\pm}}\cos2k_{R}^{\pm}x = 2\mathcal{N}_{G}^2\cos2k_{R}^{\pm}x\,.
\end{equation}
As this derivation shows, the result for the difference is quite general (up to a normalization factor), given that the form factor has only the kinematic pole $\sim(Q^2-4k_{R}^2)^{-1}$ (note that the poles at $Q=\pm2\xi_{R}$ cancel out thanks to the conditions (\ref{eq:polcondplus}), (\ref{eq:polcondminus})). We also note that the difference can be written as a sum of squares of {\em free\,} solutions of $\pm$ parity, {\em viz}
\begin{displaymath}
2\cos2kx = \left\lbrack\frac{e^{ikx}+e^{-ikx}}{\sqrt{2}}\right\rbrack^2 + \left\lbrack\frac{e^{ikx}-e^{-ikx}}{\sqrt{2}}\right\rbrack^2\,,
\end{displaymath}
so that $(\psi_{G}^{\pm}(x))^2$ and $D_{R}^{(\pm)}(x)$ encode the same non-trivial information about the dynamics of the resonance formation. A possible drawback of the use of $D_{R}^{(\pm)}(x)$ is that $D_{R}^{(-)}(0)\not=0$, while $\psi_{G}^{-}(0)=0$ in accord with the odd parity of the state. However, the advantage of $D_{R}^{(\pm)}(x)$ is that, given those densities, one can directly compute the according form factors as one does it, e.g., for a charge distribution in electrodynamics, and obtain approximations for measurable amplitudes like $\mathcal{M}(\pm k,k;Q,E)$ in the resonance region via relations like Eq.~(\ref{eq:defFR}). It is also worth mentioning that the integrals $\int dx\,x^2D_{R}^{(\pm)}(x)$ yield the analytic continuation of the squared radii of Eq.~(\ref{eq:xsqrB}) to the unphysical Riemann sheet.\\
Moreover, one can consider certain limiting cases of infinitely strong potentials, where some resonances can become stable states (an example is provided in Sec.~IV of \cite{Bruns:2018huz}). Increasing the potential towards this limit, the spatial densities of the stable states are approached smoothly by the normalized densities $D_{R}^{(\pm)}(x)$ (compare e.g. Figs.~1 and 2 in \cite{Bruns:2018huz}), while the corresponding functions $\psi_{G}^{\pm}(x)$ are non-normalizable for any finite value of the potential strength.\\

\begin{figure}[h]
\centering
\subfigure[\,]{\includegraphics[width=0.45\textwidth]{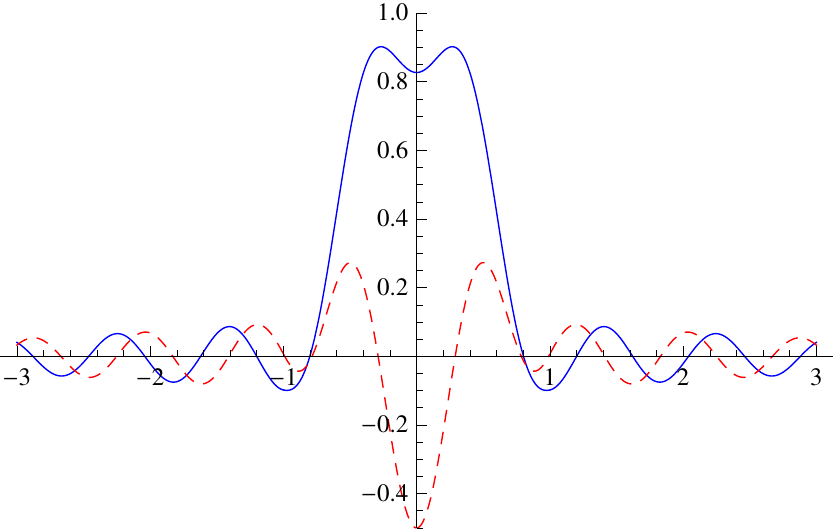}}
\subfigure[\,]{\includegraphics[width=0.45\textwidth]{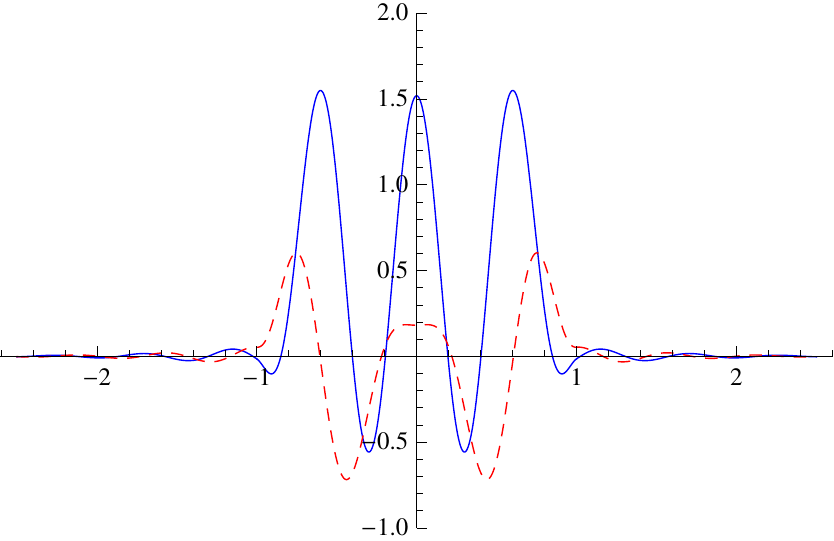}}
\caption{Real (blue) and imaginary (red, dashed) part of the density functions $D_{R}^{(+)}(x)$\, for the two lowest-lying ``$+$'' resonances for $\mu=2,\,d=1$, $V_{0}=3$. The poles are located at $k_{R,a}^{+}=3.740-0.161\,i$ and $k_{R,b}^{+}=5.657-0.876\,i$, respectively.}
\label{fig:sRplus}
\end{figure}

\begin{figure}[h]
\centering
\includegraphics[width=0.65\textwidth]{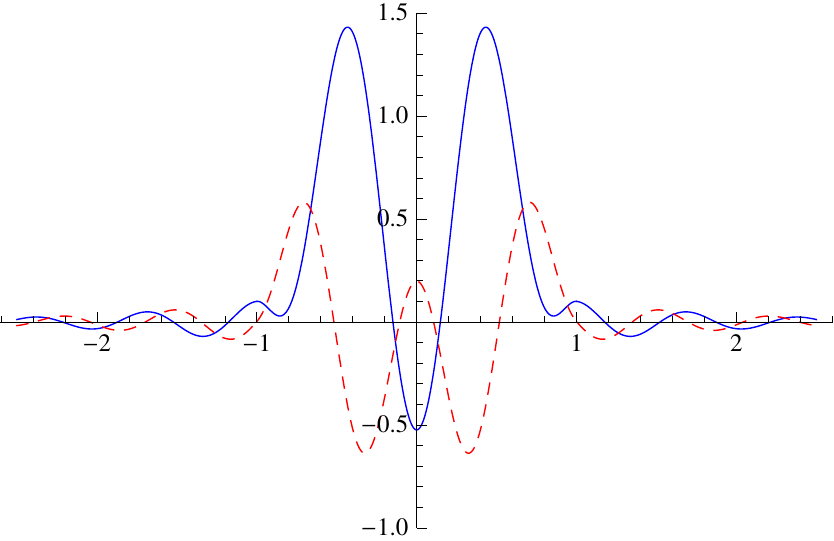}
\caption{Real (blue) and imaginary (red, dashed) part of the density function $D_{R}^{(-)}(x)$\, for the lowest-lying ``$-$'' resonance for $\mu=2,\,d=1$, $V_{0}=3$. The pole is located at $k_{R}^{-}=4.524-0.516\,i$.}
\label{fig:sRminus}
\end{figure}

Here are some examples for the densities discussed above, where we always set $d=1,\,\mu=2$. We start with a {\em repulsive\,} well (or square wall, $V_{0}=+3$). The densities $D_{R}(x)$ for the two lowest resonances of positive parity are plotted in Fig.~\ref{fig:sRplus}, the one for the lowest resonance of negative parity in Fig.~\ref{fig:sRminus}\,.


\begin{figure}[h]
\centering
\subfigure[\,$|\psi_{B}^{+}(x)|^2$]{\includegraphics[width=0.45\textwidth]{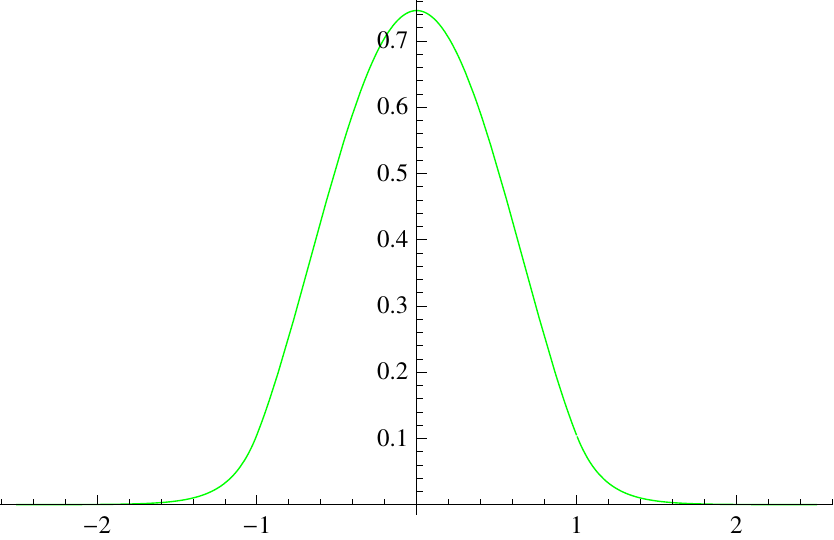}}
\subfigure[\,$|\psi_{B}^{-}(x)|^2$]{\includegraphics[width=0.45\textwidth]{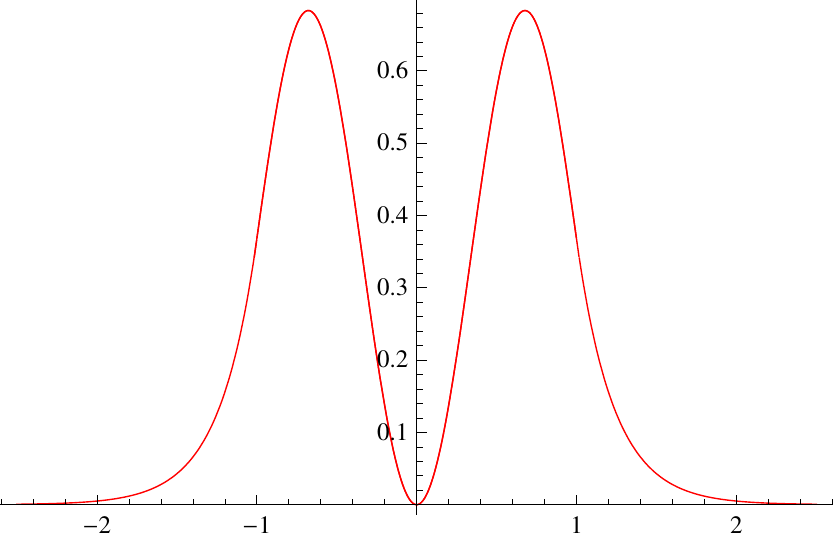}}
\caption{Probability densities for the ``$\pm$'' bound states at $\kappa_{B}^{+}=2.931$ and $\kappa_{B}^{-}=2.150$, respectively. Here, $V_{0}=-2.5$.}
\label{fig:Bprobdens}
\end{figure}

For an attractive well, there are also bound states. We show corresponding probability densities for $V_{0}=-2.5$ in Fig.~\ref{fig:Bprobdens}\,, so that the reader can compare the form of the graphs for those densities with Figs.~\ref{fig:sRplus} and \ref{fig:sRminus}\,.

\begin{figure}[h]
\centering
\subfigure[\,$D_{R}^{(+)}(x)$]{\includegraphics[width=0.45\textwidth]{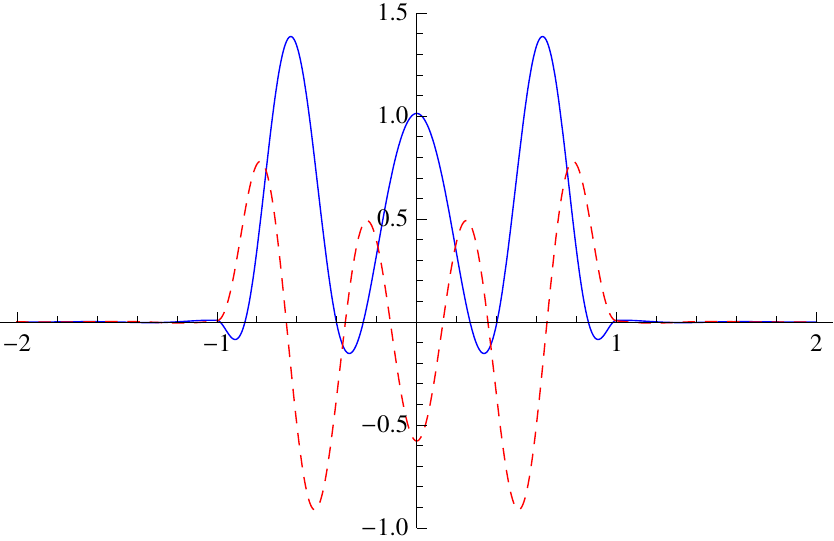}}
\subfigure[\,$|D_{R}^{(+)}(x)|$]{\includegraphics[width=0.45\textwidth]{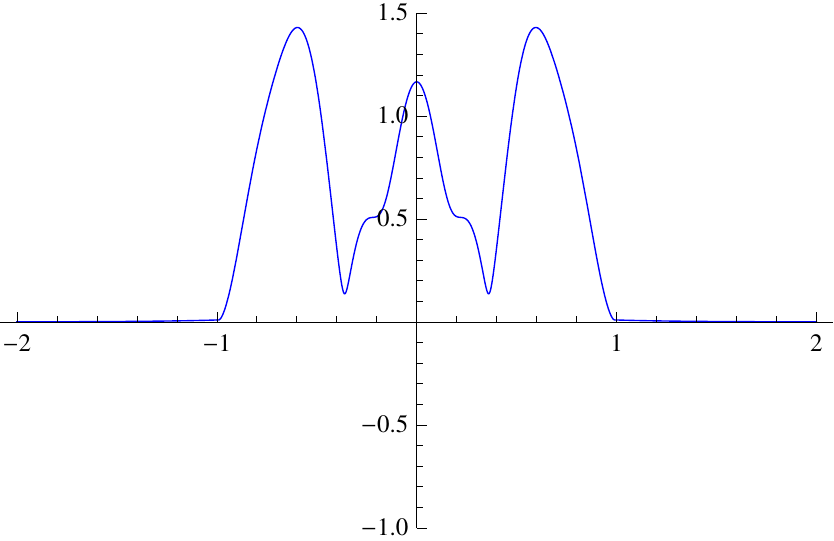}}
\caption{Real (blue) and imaginary (red, dashed) part of the density function $D_{R}^{(+)}(x)$\, for the lowest-lying ``$+$'' resonance ($V_{0}=-2.5$). The pole is located at $k_{R}^{+}=5.201-1.507\,i$. In the right-hand plot, we show the modulus of the density function.}
\label{fig:bRplus}
\end{figure}

For the same value of $V_{0}$, we find a very broad resonance in the positive-parity sector, which has a width comparable to its ``mass'', $E_{R}^{+}=6.195-3.920\,i$. We do not expect a close similarity with the structure of a bound state for this case, and approximations based on the pertaining resonance density are not expected to work well. We show this density nonetheless in Fig.~\ref{fig:bRplus}.

\clearpage
\section{Conclusion and outlook}
\label{sec:conclusio}

We have solved the Lippmann-Schwinger equation for the simple case of a square well potential in one space dimension. The form factors associated with the probability densities of the bound states in the well can be either computed from the knowledge of the bound-state wave functions (which can be obtained with the methods of elementary-school quantum theory), or from the off-shell solution $\mathcal{T}(q',q;E)$ of the LSE, see Eq.~(\ref{eq:FFgeneral}). The second method offers a straightforward generalization to the case of resonances, employing analytic continuation arguments for amplitudes like $\mathcal{M}(q',q;Q,E)$ of Eqs.~(\ref{eq:Mfull}), (\ref{eq:Mdp}) (for fixed momentum transfer $Q$ and on-shell momenta $q',q$). Applying inverse Fourier transformations in $Q$ to the extracted resonance form factors, one finds that the bound-state probability densities appearing on the r.h.s. of Eq.~(\ref{eq:psiBcheck}) are replaced by ``resonance densities'' $D_{R}(x)$, which are themselves related in a simple way to the (non-normalizable) ``resonance wave functions'' $\psi_{G}(x)$. We think that our derivations for the case of the square well make it evident that very similar methods and results would also apply for other potentials of finite range (compare also the results of \cite{Bruns:2018huz}). \\
Presumably, none of the findings presented here will be ``big news'' for the experts in the field of resonance physics. However, it is often valuable to have an exactly solvable toy model at hand, to test certain assumptions made in the analysis of more complicated problems, and to gain some general intuition. In this regard, we hope that our study might contribute to a better understanding of the physics of resonances in more realistic situations (besides possibly serving a pedagogical purpose). In hadron physics, e.g., many complications will arise: the resonances can in general decay into a lot of different multi-particle channels, there are various form factors for all kinds of associated ``charges'', relativistic dynamics will usually have to be accounted for, and one might prefer to work with light-cone wave functions or distribution amplitudes, etc. Maybe the article at hand can serve as a starting point for such more involved studies. 

\section*{Acknowledgement}
This work was supported by the Czech Science Foundation GACR grant 19-19640S.

\clearpage
\begin{appendix}

\section{Wave functions for the square well potential}
\label{app:LSpotwellwf}
\def\theequation{\Alph{section}.\arabic{equation}}
\setcounter{equation}{0}
Bound-state wave functions of even $(+)$ and odd $(-)$ parity for the potential well, pertaining to an energy eigenvalue $E_{B}$ with $V_{0}<E_{B}<0$, can be written as
\begin{eqnarray}
  \psi_{B}^{+}(x) &=& \mathcal{N}_{B}\biggl(\theta(-d-x)e^{\kappa x} + \theta(d-x)\theta(x+d)e^{-\kappa d}\frac{\cos\xi x}{\cos\xi d} + \theta(x-d)e^{-\kappa x}\biggr)\,,\label{eq:psiBp}\\
  \psi_{B}^{-}(x) &=& \mathcal{N}_{B}\biggl(\theta(-d-x)e^{\kappa x} - \theta(d-x)\theta(x+d)e^{-\kappa d}\frac{\sin\xi x}{\sin\xi d} - \theta(x-d)e^{-\kappa x}\biggr)\,,\label{eq:psiBm}\\
  \mathcal{N}_{B} &=& \frac{\xi\sqrt{\kappa}e^{\kappa d}}{\sqrt{(1+\kappa d)(\kappa^2+\xi^2)}}\,,\quad \kappa = \sqrt{2\mu(-E_{B})}>0\,,\quad \xi = \sqrt{2\mu(E_{B}-V_{0})}>0\,.\label{eq:psiBadd}
\end{eqnarray}
Here $(\kappa=-ik,\xi)$ are positive solutions of Eq.~(\ref{eq:polcondplus}) or (\ref{eq:polcondminus}), respectively. In the following, these solutions will be denoted by $\kappa_{B}^{\pm}$ and $\xi_{B}^{\pm}$. The momentum-space wave functions are
\begin{equation}
  \tilde{\psi}_{B}^{+}(p) = \int_{-\infty}^{\infty}dx\,e^{-ipx}\psi_{B}^{+}(x)\,,\qquad \tilde{\psi}_{B}^{-}(p) = \int_{-\infty}^{\infty}dx\,e^{-ipx}\psi_{B}^{-}(x)\,.
\end{equation}
These wave functions satisfy the Schr\"odinger equation in the form
\begin{eqnarray}
  \left(\frac{p^2}{2\mu}-E_{B}\right)\tilde{\psi}_{B}^{+}(p) &=& -\int_{-\infty}^{\infty}dx\,e^{-ipx}V(x)\psi_{B}^{+}(x) \nonumber \\ &=& \frac{2V_{0}\mathcal{N}_{B}e^{-\kappa_{B}^{+} d}}{p^2-(\xi_{B}^{+})^2}\left(\kappa_{B}^{+}\cos(pd)-p\sin(pd)\right)\,,\label{eq:SchrBpotplus}\\
  \left(\frac{p^2}{2\mu}-E_{B}\right)\tilde{\psi}_{B}^{-}(p) &=& -\int_{-\infty}^{\infty}dx\,e^{-ipx}V(x)\psi_{B}^{-}(x) \nonumber \\ &=& \frac{2iV_{0}\mathcal{N}_{B}e^{-\kappa_{B}^{-} d}}{p^2-(\xi_{B}^{-})^2}\left(\kappa_{B}^{-}\sin(pd)+p\cos(pd)\right)\,.\label{eq:SchrBpotminus}
\end{eqnarray}
The above expressions are non-singular at $p=\pm\xi^{\pm}_{B}$ due to the conditions Eqs.~(\ref{eq:polcondplus}), (\ref{eq:polcondminus}). We also note that the r.h.s. of the last two equations tend to $\pm\frac{\kappa^{\pm}_{B}}{\mu}\mathcal{N}_{B}$ if $p\rightarrow k_{B}^{\pm}\equiv i\kappa_{B}^{\pm}$.\\
The probability to find the bound particle outside the range of the potential is given by
\begin{equation}
P(|x|>d) = \int_{-\infty}^{-d}dx\,\psi_{B}^{\pm\ast}(x)\psi_{B}^{\pm}(x) + \int_{d}^{\infty}dx\,\psi_{B}^{\pm\ast}(x)\psi_{B}^{\pm}(x) = \frac{\mathcal{N}_{B}^2}{\kappa^{\pm}_{B}}e^{-2\kappa^{\pm}_{B}d}\,.
\end{equation}
Up to a different normalization factor $\mathcal{N}_{B}$, this probability will be of the same form for any finite-range potential ($0<d<\infty$). The splitting $1=P(|x|>d)+P(|x|<d)$ agrees with the splitting into ``compositeness'' (first term) and ``elementariness'' (second term) \cite{Weinberg:1965zz,Baru:2003qq,Aceti:2012dd,Sekihara:2014kya} up to the exponential factor $e^{-2\kappa^{\pm}_{B}d}$, which, however, tends to one as $\kappa^{\pm}_{B}d\rightarrow 0$ (but, dropping this exponential, it is not necessarily true that the ``compositeness'' $\mathcal{N}_{B}^2/\kappa^{\pm}_{B}\leq 1$).
\newpage
Tentatively applying these remarks concerning our simple one-dimensional framework to the case of the deuteron, assuming that the binding force between the proton and the neutron has a finite range, one would interpret the results of \cite{Weinberg:1965zz} as meaning that the chance to find the proton and the neutron separated by a distance greater than the interaction range is $\gtrsim 80\%$. \\
\quad \\
For the sake of completeness, we also treat the continuum solutions ($E>0$, $k=+\sqrt{2\mu E}>0$, $\xi=\sqrt{2\mu(E-V_{0})}$).
To this end, we define transmission and reflection amplitudes $\tau$ and $\rho$\,,
\begin{equation}
\tau(E) = 1-\frac{2\pi i\mu}{k}\mathcal{T}(+k,k;E)\,,\qquad \rho(E) = -\frac{2\pi i\mu}{k}\mathcal{T}(-k,k;E)\,,\label{eq:taurhopot}
\end{equation}
and note the relations $\tau+\rho-1=2\left(-\frac{2\pi i\mu}{k}\right)\mathcal{T}_{\mathrm{s}}\,$, $\,\tau-\rho-1=2\left(-\frac{2\pi i\mu}{k}\right)\mathcal{T}_{\mathrm{p}}\,$. Explicitly,
\begin{equation}\label{eq:taurhopotexplicit}
  \tau(E) =\frac{4k\xi\,e^{2i(\xi-k)d}}{(k+\xi)^2-(k-\xi)^2e^{4i\xi d}}\,,\quad \rho(E)=\frac{(k^2-\xi^2)e^{-2ikd}(1-e^{4i\xi d})}{(k+\xi)^2-(k-\xi)^2e^{4i\xi d}}\,.
\end{equation}
Unitarity of the on-shell scattering amplitude can be expressed in the form $|\tau(E)\pm\rho(E)|=1$ for $E>0$ (``every incoming particle is either reflected or transmitted'').\\
Solutions of type $L$ (``incoming particle from the left'') can be written as
\begin{eqnarray}
  \psi_{E}^{L}(x) &=& \frac{1}{\sqrt{2\pi}}\biggl(\theta(-d-x)\left(e^{ikx}+\rho(E)e^{-ikx}\right) + \theta(d-x)\theta(x+d)\left(a(E)e^{i\xi x}+b(E)e^{-i\xi x}\right) \nonumber \\
  & & \hspace{6.45cm} +\,  \theta(x-d)\tau(E)e^{ikx}\biggr)\,,\label{eq:psiLpot} \\
  a(E) &=& \frac{e^{i(\xi+k)d}\tau(E)-e^{-i(\xi-k)d}\rho(E)-e^{-i(\xi+k)d}}{2i\sin2\xi d}\,,\nonumber \\
  b(E) &=& \frac{e^{i(\xi-k)d}-e^{-i(\xi-k)d}\tau(E)+e^{i(\xi+k)d}\rho(E)}{2i\sin2\xi d}\,.\quad \nonumber
\end{eqnarray}
Solutions of type $R$ (``incoming particle from the right''):
\begin{eqnarray}
\psi_{E}^{R}(x) &=& \frac{1}{\sqrt{2\pi}}\biggl(\theta(-d-x)\tau(E)e^{-ikx} + \theta(d-x)\theta(x+d)\left(a(E)e^{-i\xi x}+b(E)e^{i\xi x}\right) \nonumber \\
  & & \hspace{6.45cm} +\,  \theta(x-d)\left(e^{-ikx}+\rho(E)e^{ikx}\right)\biggr)\,\label{eq:psiRpot}
\end{eqnarray}
Continuum solutions of even and odd parity can now be constructed as
\begin{equation}\label{eq:psipmpot}
\psi_{E}^{+}(x) = \frac{1}{\sqrt{2}}(\psi_{E}^{L}(x)+\psi_{E}^{R}(x))\,,\qquad \psi_{E}^{-}(x) = \frac{1}{\sqrt{2}}(\psi_{E}^{L}(x)-\psi_{E}^{R}(x))\,.
\end{equation}

\newpage
\section{Results for $\mathcal{M}_{\mathrm{d.p.}}$}
\label{app:Mdp}
\def\theequation{\Alph{section}.\arabic{equation}}
\setcounter{equation}{0}

In this appendix we give the explicit expressions for $\mathcal{M}_{\mathrm{d.p.}}^{\mathrm{fw}}:=\mathcal{M}_{\mathrm{d.p.}}(k,k;Q,E)$ and $\mathcal{M}_{\mathrm{d.p.}}^{\mathrm{bw}}:=\mathcal{M}_{\mathrm{d.p.}}(-k,k;Q,E)$, evaluated on the physical Riemann sheet ($\mathrm{Im}\,k>0$).
\begin{eqnarray*}
  \mathcal{M}_{\mathrm{d.p.}}^{\mathrm{fw}/\mathrm{bw}} &=& \frac{8k(k^2-\xi^2)e^{4i\xi d}\left(\mathcal{M}_{\mathrm{cos}}^{\mathrm{fw}/\mathrm{bw}}\cos(Qd) + \frac{\sin(Qd)}{Q}\mathcal{M}_{\mathrm{sin}}^{\mathrm{fw}/\mathrm{bw}}\right)}{\pi(Q^2-4\xi^2)(4k^2-Q^2)((k+\xi)^2-Q^2)((k-\xi)^2-Q^2)((k+\xi)^2-e^{4i\xi d}(k-\xi)^2)^2}\,,\\
  \mathcal{M}_{\mathrm{cos}}^{\mathrm{fw}} &=& (Q^2-4\xi^2)\biggl(2ik\xi(4k^2-Q^2)\cos(2\xi d) + \left(3k^2(k^2-Q^2)+(6k^2+Q^2)\xi^2-\xi^4\right)\sin(2\xi d)\biggr)f_{\mathrm{p}}\\
  &+& 2k e^{-2ikd}\biggl(2ik\xi^2(4k^2-Q^2)(4\xi^2-Q^2)\cos(2\xi d) \\ &+& \left(4k^6-12k^4Q^2+5k^2Q^4+(4k^4-8k^2Q^2-3Q^4)\xi^2+4(7k^2+Q^2)\xi^4-4\xi^6\right)\xi\sin(2\xi d)\biggr)\,,\\
  \mathcal{M}_{\mathrm{sin}}^{\mathrm{fw}} &=& (Q^2-4\xi^2)\biggl((4k^2-Q^2)(k^2+Q^2-\xi^2)\xi\cos(2\xi d)\\  & & \hspace{6.5cm} -ik(2k^4-k^2Q^2-Q^4+5Q^2\xi^2-2\xi^4)\sin(2\xi d)\biggr)f_{\mathrm{p}}\\ &+& e^{-2ikd}\biggl(-2k\xi^2(4k^2-Q^2)(2k^4-5k^2Q^2+Q^4+Q^2\xi^2-2\xi^4)\cos(2\xi d) \\ &+&  i(4\xi^2-Q^2)\left(8k^6-5k^4Q^2+3k^2Q^4-(8k^4+4k^2Q^2+Q^4)\xi^2+Q^2\xi^4\right)\xi\sin(2\xi d)\biggr)\,,\\
  \mathcal{M}_{\mathrm{cos}}^{\mathrm{bw}} &=& -i(2k-Q)(Q^2-4\xi^2)((k-Q)^2-\xi^2)\xi f_{\mathrm{p}} \\
  &+& \frac{i}{2}e^{-2ikd}((k-Q)^2-\xi^2)\biggl(-k^3Q(k+Q)(4k+Q) + k(4k^3+20k^2Q+14kQ^2-3Q^3)\xi^2 \\ & & \hspace{4.4cm} - (40k^2+Q^2)\xi^4 + 4\xi^6 \\
  &+& (k^2-\xi^2)\left(kQ(k+Q)(4k+Q)-(4k^2+16kQ+Q^2)\xi^2+4\xi^4\right)\cos(4\xi d) \\
  &+& (k^2-\xi^2)(2k-Q)(4k^2+6kQ+Q^2-4\xi^2)i\xi\sin(4\xi d)\biggr)\,,\\
  \mathcal{M}_{\mathrm{sin}}^{\mathrm{bw}} &=& (2k-Q)Q(Q^2-4\xi^2)((k-Q)^2-\xi^2)\xi f_{\mathrm{p}} \\
  &+& \frac{1}{2}e^{-2ikd}((k-Q)^2-\xi^2)\biggl(-k^3Q^2(k+Q)(4k+Q) \\ & & \hspace{4.4cm} + \,k(32k^4+36k^3Q-4k^2Q^2-2kQ^3+5Q^4)\xi^2\\
  & & \hspace{4.4cm} -(32k^3+8k^2Q+8kQ^2+Q^3)\xi^4 + 4Q\xi^6 \\
  &+& (k^2-\xi^2)Q\left(kQ(k+Q)(4k+Q)-(4k^2+16kQ+Q^2)\xi^2+4\xi^4\right)\cos(4\xi d) \\
  &+& (k^2-\xi^2)Q(2k-Q)(4k^2+6kQ+Q^2-4\xi^2)i\xi\sin(4\xi d)\biggr)\,,\\
  f_{\mathrm{p}} &:=& 2k\xi\cos(2\xi d)-i(k^2+\xi^2)\sin(2\xi d)\,.
\end{eqnarray*}
Note that $f_{\mathrm{p}}$ is odd in $\xi$, and vanishes at the poles of the on-shell scattering amplitude.
\end{appendix}

\end{document}